\documentclass[reprint, amsmath, amssymb, aps, floatfix]{revtex4-2}

\usepackage{color}
\usepackage[dvipsnames]{xcolor}
\usepackage[normalem]{ulem}
\usepackage{float}

\usepackage{graphicx}
\usepackage{dcolumn}
\usepackage{bm}
\usepackage{braket}
\usepackage{amsmath}

\begin{document}

\preprint{APS/123-QED}

\author{Joseph G. Smith$^{1,2}$}
 \email{jgs46@cam.ac.uk}
\author{Crispin H. W. Barnes$^1$}
 \email{chwb101@cam.ac.uk}
\author{David R. M. Arvidsson-Shukur$^{2}$}
 \email{drma2@cam.ac.uk}
\affiliation{$^1$Cavendish Laboratory$,$ Department of Physics$,$  University  of  Cambridge$,$  Cambridge$,$  CB3  0HE$,$  United  Kingdom}
\affiliation{$^2$Hitachi Cambridge Laboratory$,$ J. J. Thomson Ave.$,$ Cambridge$,$ CB3 0HE$,$ United Kingdom}

\title{An iterative quantum-phase-estimation protocol for near-term quantum hardware}

\date{\today}

\begin{abstract}

{Given $N_{\textrm{tot}}$ applications of a unitary operation with an unknown phase $\theta$, a large-scale fault-tolerant quantum system can {reduce} an estimate's {error} scaling from $\mathcal{O} \left[ 1 / \sqrt{N_{\textrm{tot}}} \right]$ to $\mathcal{O} \left[ 1 / {N_{\textrm{tot}}} \right]$. Owing to the limited resources available to near-term quantum devices, entanglement-free protocols have been developed, which achieve a $\mathcal{O} \left[ \log(N_{\textrm{tot}}) / N_{\textrm{tot}} \right]$ {mean-absolute-error} scaling. Here, we propose a new two-step protocol for near-term phase estimation, with an improved {error} scaling. Our protocol's first step produces several low-{standard-deviation} estimates of $\theta $, within $\theta$'s parameter range. The second step iteratively hones in on one of these estimates. Our protocol's {mean absolute error} scales as $\mathcal{O} \left[ \sqrt{\log (\log N_{\textrm{tot}})} / N_{\textrm{tot}} \right]$. Furthermore, we demonstrate a reduction in the constant scaling factor and the required circuit depths: our protocol can outperform the asymptotically optimal quantum-phase estimation algorithm for realistic values of $N_{\textrm{tot}}$.}

\end{abstract}

\maketitle

\textit{Introduction}.---The task of finding an unknown parameter $\theta$ of a unitary operation $\hat{U}(\theta)$ requires \textit{phase estimation}, one of the most prominent tasks in quantum-information processing. Various forms of phase estimation occur in, for example: the subroutines of quantum algorithms \cite{Qcount, Shor, david, linear_eqs}; protocols to find ground-state energies \cite{VQE}; gravitational-wave detection \cite{LIGO}; fixed-reference-frame sharing \cite{RG}; synchronization of clocks \cite{Burgh}; and, famously, in the measurement of time \cite{clocks}. To measure an unknown quantity of interest, $\theta$, a quantum probe $\psi_0$ is subjected to the unitary operation $\hat{U}(\theta)$, such that the output probe $\psi_{\theta}$ carries useful information \cite{Lloyd1}. This information is then accessed via measurements. As quantum measurements are probabilistic in nature, statistics lead to a bound on the {error} of any estimate of $\theta$, $\widetilde{\theta}$. (Throughout this manuscript, estimates of the quantity $X$ are distinguished using $\widetilde{X}$). By utilizing quantum phenomena, these bounds can be improved. In particular, if $\hat{U}(\theta)$ is queried $N_{\textrm{tot}}$ times, each time using a separate probe, the {error}, $\Delta \tilde{\theta}$, scales asymptotically to the shot-noise limit $\Delta \tilde{\theta} \propto 1 / \sqrt{N_{\textrm{tot}}}$. Using quantum coherence or entanglement, the scaling can be improved to the Heisenberg Limit: $\Delta \tilde{\theta} \propto 1 / N_{\textrm{tot}}$ \cite{intfer_2, metro_1, HL2}. The ability to decrease the {error} in this way constitutes one of the most tractable technological applications for quantum advantage. 

An example of an algorithm that achieves the Heisenberg Limit is the quantum-phase-estimation (QPE) algorithm. This algorithm uses the inverse Fourier transform on a set of entangled probes to provide an estimate of a phase \cite{algorithms, nielsen}. However, the circuit depths, coherence times and gate fidelities needed for practical use of this algorithm are far beyond the realistic regime of noisy intermediate-scale quantum devices \cite{QPE_limitations}. Instead, one can use Maximum Likelihood Estimators (MLEs) to analyse the measurement outcomes of a single, shallower circuit \cite{QFT_vs_MLE}. These quantum-classical strategies involve quantum-probe preparation followed by ``classical'' measurements, which sample individual probes separately \cite{Lloyd2}. A significant, but often overlooked, drawback of MLE strategies that sample only one circuit, is that their {error} minimization  leads to a \textit{point unidentified} estimate $\tilde{\theta}$ (see below). That is, the MLE cannot distinguish between several possible values of $\theta$ \cite{set_id, set_id2, set_id3}. To combat this, MLE-based protocols have been introduced, which iteratively measure multiple circuits \cite{loophole, program} to avoid point unidentification. To our knowledge, until now, the best MLE-based protocol achieves a {mean-absolute-error} scaling of $\Delta \tilde{\theta} \propto \log{(N_{\textrm{tot}})} / N_{\textrm{tot}}$ \cite{RG, Burgh}. 

In this Letter, we construct a two-step protocol that splits the phase estimation problem into a quantum-classical strategy and a point-identification strategy. Our protocol, which does not suffer from point unidentification, achieves lower {mean absolute errors}, and has shallower circuits, than existing phase-estimation protocols. When the point identification is conducted iteratively, our protocol achieves an {mean absolute error} scaling of {$\Delta \tilde{\theta} \propto \sqrt{\log ( \log{N_{\textrm{tot}}})} / N_{\textrm{tot}}$. This scaling is better than previous iterative protocols.} Additionally, we show that our protocol, which requires no entanglement between probes, achieves estimates with a lower {error} than those acquired by the QPE algorithm, for experimentally realistic circuit depths and values of $N_{\textrm{tot}}$.

\begin{figure}
    \centering
    \includegraphics[width = 8.6cm]{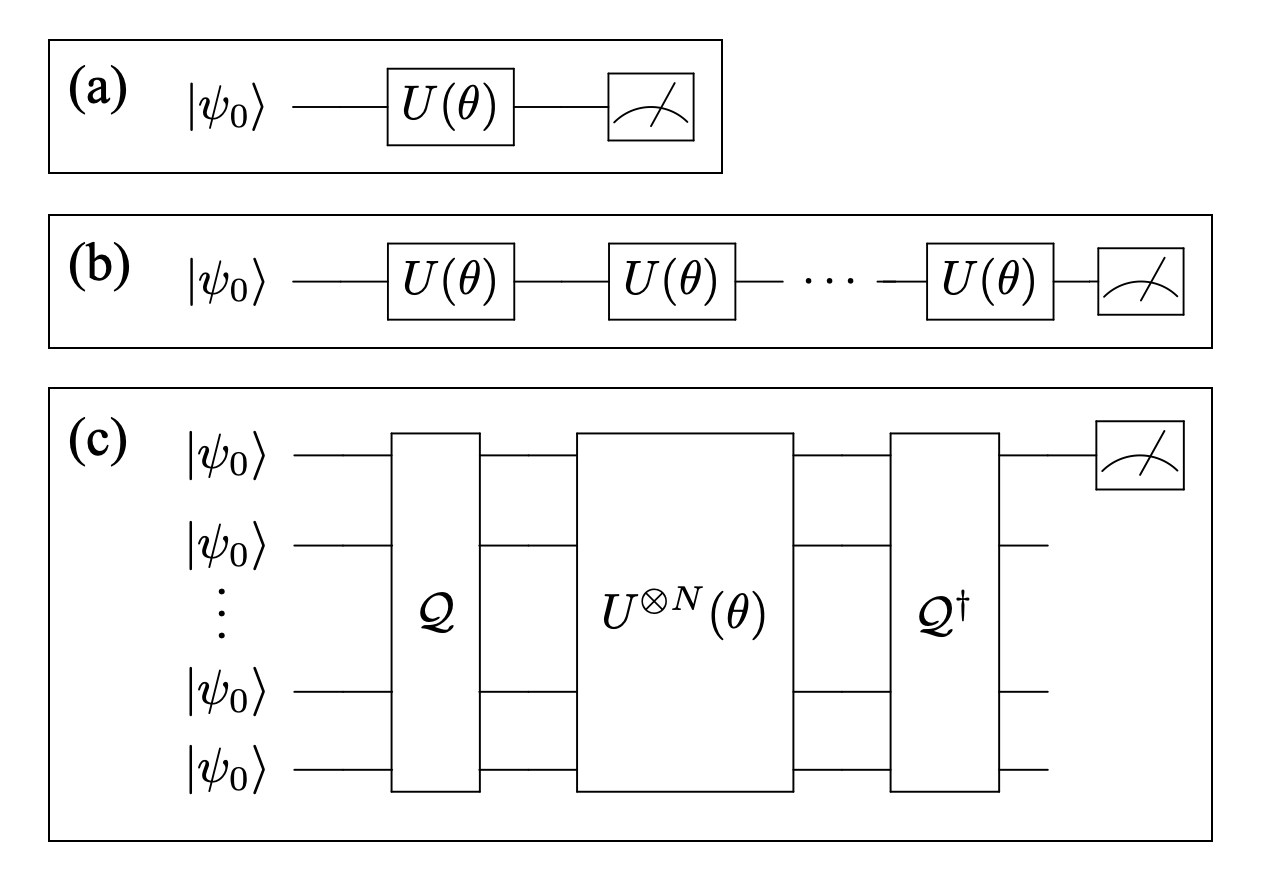}
    \caption{Quantum circuits used to estimate $\theta$ with \textbf{(a)} one application of $\hat{U}(\theta)$ and \textbf{(b)} $N$ coherent applications of $\hat{U}(\theta)$ in series. \textbf{(c)} Phase estimation via entanglement of $N$ probes. The gate $\mathcal{Q}$ is used to entangle the probes into the GHZ state from an initial state $\ket{\psi_0}$.}
    \label{fig:1_circuit}
\end{figure}

\textit{Background}.---Throughout this work, we focus on Stone's encoded unitaries with a fixed $\theta$ \cite{Stone}: $\hat{U}(\theta)=e^{i\theta\hat{A}}$. $\hat{A}$ is a Hermitian generator independent of $\theta$ \footnote{Unitaries where $\hat{A}$ has explicit $\theta$ dependence can often be recast as $e^{i \theta^\prime \hat{A}^\prime}$ such that $\hat{A}^\prime$ has no $\theta$ dependence.}.
We also focus on  optimal phase-estimation, by setting the input probe states to $\ket{\psi_0} = \frac{1}{\sqrt{2}} \left( \ket{a_{\textrm{min}}} + \ket{a_{\textrm{max}}} \right)$, where $\ket{a_{\textrm{min}}}$ and $\ket{a_{\textrm{max}}}$ are  eigenstates corresponding to minimum and maximum eigenvalues of $\hat{A}$, respectively. This state maximizes the acquired phase difference from  $\hat{U}(\theta)$ \cite{Lloyd2}. After suitable parameter rescaling, we can write  (ignoring a global phase) the unitary operation as
\begin{equation}
\label{eqn:U}
    \hat{U}(\theta) = e^{-i\theta/2}\ket{a_{\textrm{min}}} \!\! \bra{a_{\textrm{min}}}+e^{i\theta/2}\ket{a_{\textrm{max}}} \!\! \bra{a_{\textrm{max}}},
\end{equation}
where $\theta \in [0, 2\pi)$. Applying $\hat{U}(\theta)$ sequentially $N$ times is equivalent to applying $\hat{U}(N\theta)$ once. The probability that the probe remains in the state $\ket{\psi_0}$ after $N$ applications of $\hat{U}(\theta)$ is
\begin{equation}
    p_0(N, \theta) = |\!\bra{\psi_0} \! \hat{U}^N(\theta) \! \ket{\psi_0}\!|^2 = \frac{1}{2}\left[1 + \cos (N \theta)\right].
    \label{eqn:p_0(N)}
\end{equation}
Alternatively, one could prepare $N$ probes in a GHZ state and apply $\hat{U}(\theta)$ once to each probe in parallel \cite{Lloyd1}. [See Fig. \ref{fig:1_circuit}(b) and (c).] It is possible to estimate $\theta$ through an estimate of $p_0(N,\theta)$: 
\begin{equation}
    \theta = \pm \frac{1}{N} \arccos{\left[ 2p_0(N,\theta) -1 \right]} + \frac{2\pi l}{N} ,
    \label{eqn:theta}
\end{equation}
for integer $l$. The estimate of $p_0(N,\theta)$ can be achieved by first preparing $\nu$ probes in state $\ket{\psi_0}$, then applying a $\hat{U}(\theta)$ operation $N$ times to each probe, and finally measuring the probes in the $\{ \ket{\psi_0}, \ket{\psi_0}^{\bot}\}$ basis. If $x$ of these $\nu$ measurements correspond to the $\ket{\psi_0}$ outcome, MLEs \cite{MLE} can be used to estimate $p_0(N,\theta)$: $\widetilde{p}_0(N,\theta)=\frac{x}{\nu}$. The associated {standard deviation is $\sigma_{\widetilde{p}_0(N,\theta)} \geq \sqrt \frac{p_0(N,\theta) \left(1 - p_0(N,\theta) \right)}{\nu}$} \cite{FI_tutorial}. {(We distinguish the mean absolute error of an estimate of $X$, $\Delta \widetilde{X}=\mathbb{E} \left[ |X-\widetilde{X}| \right]$, from the standard deviation $\sigma_{\widetilde{X}}$ of the estimate)}. From Eq. \eqref{eqn:theta}, we see that an estimate of $\theta$ has a lower bound on the standard deviation: $ \sigma_{\widetilde{\theta}} \geq \frac{1}{N \sqrt \nu}$.
This inequality saturates for large $\nu$. The reduction in {standard deviation} by a factor $N$ arises directly from quantum coherence [in Fig. \ref{fig:1_circuit}(b)] or entanglement [in Fig. \ref{fig:1_circuit}(c)] \cite{use_ent}. Methods that do not use quantum phenomena [Fig. \ref{fig:1_circuit}(a)] have $N=1$, and achieve a {standard deviation} bounded by the Standard Quantum Limit: $\sigma_{\widetilde{\theta}} \geq \frac{1}{\sqrt \nu}$.

An obvious problem with the aforementioned quantum methods, is that for any given $p_0(N,\theta)$, $2N$ different values of $\theta \in [0,2\pi)$ satisfy Eq. \eqref{eqn:theta}. Point identification \cite{set_id, set_id2, set_id3} is needed to determine the correct $l$ and yield an unambiguous estimate of $\theta$. Even the classical method, where $N=1$, cannot distinguish between a true underlying parameter of $\theta$ or $2\pi - \theta$. In this case, one can achieve point identification by carrying out also a second circuit in which $\hat{U}(\theta)$ is followed by $\hat{U}(\pi/2)$. In the second circuit $p_0(1,\theta)$ becomes $p_0(1,\theta+\pi/2) = \frac{1}{2}\left[1-\sin(\theta) \right]$. If $p_0(1,\theta + \pi/2) < 1/2$, $\theta \in [0, \pi)$, else $\theta \in [\pi, 2\pi)$ \cite{phase_shift}. Thus, the second circuit allows us to point-identify in which subspace of the parameter range the unknown parameter lies. In the general case, $N>1$, point identification is not achieved by applying $\hat{U}(\pi/2)$ alone. One must iteratively increase $N$ and conduct corresponding quantum-classical point-identification techniques until the target $N$ is reached \cite{RG}. The point-identification procedures require measurements that do not necessarily decrease the {error} of the final estimate. Consequently, point identification leads to difficulties in reaching the Heisenberg Limit. 

Throughout this work, we take the total number of applications of $\hat{U}(\theta)$, $N_{\textrm{tot}}$, as the resource of phase-estimation protocols. That is, we compare the {error} of a protocol with $N_{\textrm{tot}}$. To investigate the viability of protocols on noisy intermediate-scale quantum hardware, we also consider the protocols' maximum circuit depth $N_{\textrm{max}}$.

\textit{Two-step protocol}.---We now introduce our protocol, which splits the phase estimation into two steps: First, a fine-tuning step that executes a circuit with $N$ applications of $\hat{U}(\theta)$ to achieve several low-standard-deviation estimates of $\theta$. Second, a point-identification step that disambiguates the estimate through either an iterative method or an application of the QPE algorithm (see below). Given a point-identification method and a value of $N_{\textrm{tot}}$, $N$ is chosen to minimize $\Delta \widetilde{\theta}$. Consider a measurement of the circuit in Fig. \ref{fig:1_circuit}(b) with $N=2^m$, where $m \in \mathbb{N}$. This corresponds to the fine-tuning step of our protocol. By defining $\theta \equiv 2\pi T$, $T \in [0, 1)$, and binary expanding $T = \sum^{\infty}_{j=1} t_{j}2^{-j}$ where $t_{j}$ is the $j^{\mathrm{th}}$ binary bit of $T$, the probability of measuring a $\ket{\psi_0}$ state, Eq. \eqref{eqn:p_0(N)}, becomes
\begin{equation}
\label{eqn:N_bits}
\begin{split}
    p_0(2^m,\theta) 
    =
    \frac{1}{2} \left[1 \pm \cos\left(\theta_{\textrm{FT}} \right)\right], 
\end{split}
\end{equation}
where $\theta_{\textrm{FT}} \equiv (2^{m} \theta  \mod \pi)  = 2\pi \sum^{\infty}_{j=m+2} t_{j}2^{m-j} $, and addition (subtraction) occurs if $t_{m+1}=0$ ($1$). We note that only the bits $t_j$ with $j>m+1$ affect $p_0(2^m,\theta)$ in this fine-tuning step.
The circuit with $N=2^m$ is executed $\nu_{\textrm{FT}}$ times and, upon counting $x_{\textrm{FT}}$ probes in the state $\ket{\psi_0}$, we estimate $\widetilde{p}_0(2^m,\theta) = \frac{x_{\textrm{FT}}}{\nu_{\textrm{FT}}}$. We then invert Eq. \eqref{eqn:N_bits} to estimate $\theta_{\textrm{FT}}$. Fine-tuning involves $\hat{U}(\theta)$ being applied $\nu_{\textrm{FT}}2^m$ times, and returns an estimate with $\sigma_{\widetilde{\theta}_{\textrm{FT}}}=\frac{1}{\sqrt{\nu_{\textrm{FT}}}}$ for large $\nu_{\textrm{FT}}$.

The next step is point identification, which involves finding the bits $t_j$ with $j\leq m+1$. These bits define the quantity $\theta_{\textrm{PI}} \equiv 2\pi \sum_{j=1}^{m+1}t_j2^{-j}$. An estimate of $\theta_{\textrm{PI}}$ can be found by a number of methods. We give two examples below. In general, this step applies $\hat{U}(\theta)$ a total of $N_{\textrm{PI}}$ times. A final estimate of $\theta$ is then given by $\widetilde{\theta}=\widetilde{\theta}_{\textrm{PI}}+2^{-m}\widetilde{\theta}_{\textrm{FT}}$ with {standard deviation} $\sigma_{\widetilde{\theta}} = 2^{-m}\sigma_{ \widetilde{\theta}_{\textrm{FT}}}$ if the point identification was successful.

If $\hat{U}(\theta)$ is applied $N_{\textrm{tot}}$ times over the two steps, $\nu_{\textrm{FT}}$ can take a maximum value of $ \lfloor 2^{-m} \left(N_{\textrm{tot}} -N_{\textrm{PI}}\right) \rfloor$. By minimizing the {standard deviation} with respect to $m$ we find an equation for $N_{\textrm{tot}}$:
\begin{equation}
    \label{eqn:min_m}
    N_{\textrm{tot}}  = N_{\textrm{PI}}  + \frac{1}{\ln2} \frac{\partial N_{\textrm{PI}}}{\partial m}.
\end{equation}
The optimal value of $m$ is the integer closest to the value of $m$ that satisfies Eq. \eqref{eqn:min_m}. Using this value, we find a bound on the {standard deviation} of the estimate:
\begin{equation}
    \sigma_{\widetilde{\theta}}\geq\sqrt{\frac{\ln2}{2^m \frac{\partial N_{\textrm{PI}}}{\partial m}}}.
    \label{eqn:cal_err}
\end{equation}
In the asymptotic limit, where $\nu_{\textrm{FT}}$ is large, this bound saturates and becomes an equality.

\begin{figure}
    \centering
    \includegraphics[width = 4.5cm]{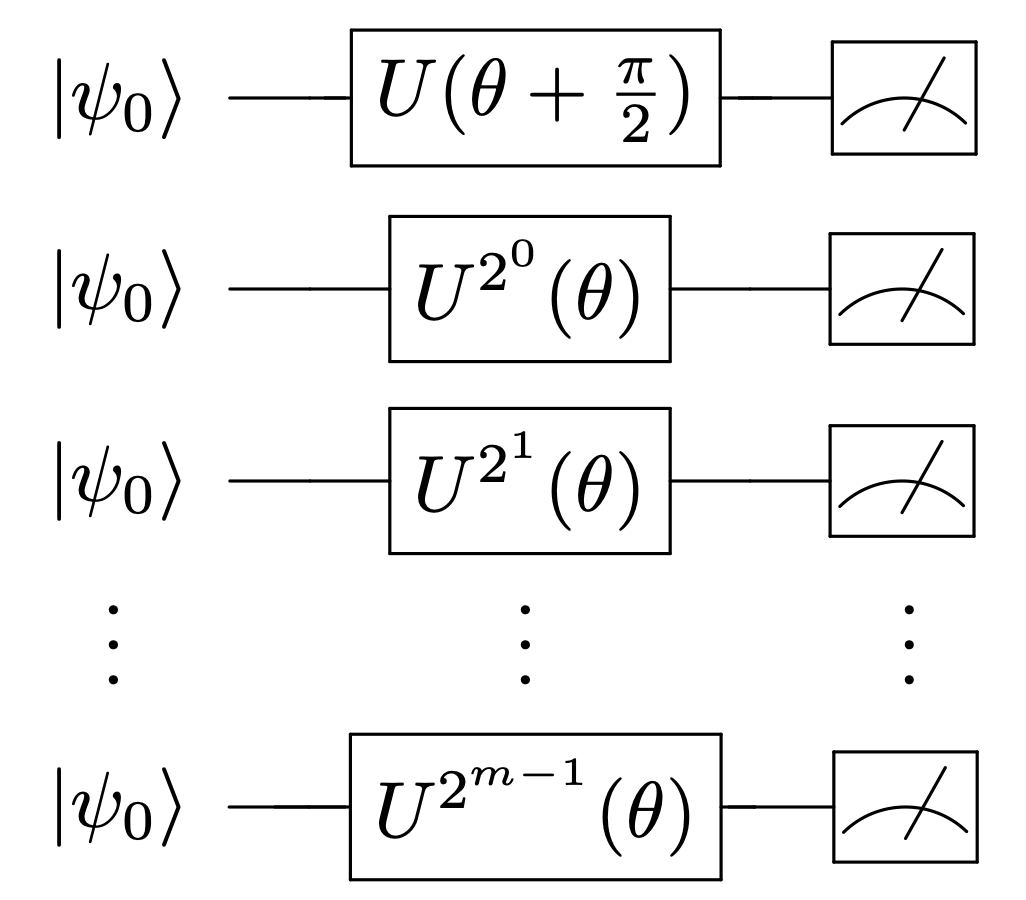}
    \caption{Circuits that are executed to estimate the first $m+1$ bits of $T$, $\theta_{\textrm{PI}}$.}
    \label{fig:it_circs}
\end{figure}

\textit{Iterative method for point identification}.---Here, we outline how to estimate $\theta_{\textrm{PI}}$ through iteration of many circuits. These circuits have varying depth, $N=2^i$, for integers $i \in [0,1,\ldots, {m-1}]$, and are executed to estimate if $p_0(2^i,\theta)>1/2$ by using the MLE method defined above. We set $m \rightarrow i$ in Eq. \eqref{eqn:N_bits}, and note that if $t_{i+2} = 1$, then $\cos \left( 2\pi \sum_{j=i+2}^\infty t_j 2^{i-j} \right) \leq 0$. Therefore, $p_0(2^i,\theta)\leq 1/2$ or $p_0(2^i,\theta)\geq 1/2$ if $t_{i+1}=0$ or $t_{i+1}=1$, respectively. If instead $t_{i+2} = 0$, the relationship between $p_0(2^i,\theta)$ and $t_{i+1}$ is the opposite. Therefore, knowing the value of the bit $t_{i+1}$ and estimating if $p_0(2^i,\theta)>1/2$ allows us to estimate the bit $t_{i+2}$. We then iterate by increasing $i$ from $0$ up to $m-1$ to estimate all of the first $m+1$ bits of $T$, bar the first bit, $t_1$. $t_1$ is estimated differently, by using  additional evolutions of $\hat{U} \left(\theta + \pi/2 \right)$ as described above. The whole iteration process is summarised in Fig. \ref{fig:it_circs}. Because the phase estimation problem is split into many circuits, this protocol allows for parallel execution.

To cap the probability, $\epsilon$, that the point-identification step fails, we need to limit the probability, $\epsilon_i$, that the $i^{\textrm{th}}$ bit of $T$ is incorrectly assigned. Thus, the circuit with $N=2^i$ phase applications must be executed a minimum number of times, $\nu_i$. A suitable $\nu_{i}$ can be calculated using the binomial distribution's Chernoff bound \cite{RG, nielsen}:
\begin{equation}
    \label{eqn:Chernoff}
    \textrm{Pr}\left[ |\widetilde{p}_0(N,\theta) - p_0(N,\theta)|  \geq \delta \right] \equiv \epsilon_i \leq 2 e^{- \nu_{i} \delta^2 / 2},
\end{equation}
where $\delta$ is the maximum allowed absolute difference between the estimated $\widetilde{p}_0(N,\theta)$ and true $p_0(N,\theta)$. Failure occurs if $\widetilde{p}_0(N,\theta)>1/2$ when $p_0(N,\theta)<1/2$ (and vice-versa): $|\widetilde{p}_0(N,\theta)-p_0(N,\theta)| \geq |\frac{1}{2} - p_0(N,\theta)|$. Hence, we choose $\delta = |\frac{1}{2} - p_0(N,\theta)|$ when solving Eq. \eqref{eqn:Chernoff}:
\begin{equation}
    \nu_{i} \geq \frac{2 \ln(2/\epsilon_i)}{(1/2 - p_0(N,\theta))^{2}} = \frac{8 \ln(2/\epsilon_i)}{\cos^{2}(N\theta)} .
\end{equation}
Problematically, one needs knowledge of $\theta$ to find  $\nu_{i}$. To resolve this, we set the denominator to a constant, $\alpha \equiv 8 \sec^{2}N\theta$, and accept that some values of $\theta$ lead to a failure probability larger than $\epsilon_i$. Increasing $\alpha$ will reduce this effect. Our protocol executes a circuit with $N=2^i$ in total $\nu_i=\alpha \ln(2/\epsilon_i)$ times. Over the whole point-identification step, we thus apply $\hat{U}(\theta)$ a number 
\begin{equation}
    N_{\textrm{PI}} = \nu_0 + \sum_{i=0}^{m-1}2^{i}\nu_i = \alpha \ln(2/\epsilon_0) + \sum_{i=0}^{m-1} \alpha 2^{i} \ln(2/\epsilon_i)
    \label{eqn:N_cal_1}
\end{equation}
times in total to estimate $\theta_{\textrm{PI}}$. We now make the assertion that the whole point-identification protocol is incorrect with maximum probability $ \epsilon$, such that $ 1-\epsilon = \prod_{i=1}^{m}(1-\epsilon_i)$. 
For small $\epsilon_i$, $\epsilon = \sum_{i=1}^{m}\epsilon_i$. We use Lagrange multipliers to minimize Eq. \eqref{eqn:N_cal_1} with this constraint. We find that $\epsilon_i=2^{i-m}\epsilon$, $\nu_i = \alpha \ln \left( 2^{m+1-i} / \epsilon\right)$ and
\begin{equation}
    N_{\textrm{PI}} =  \alpha 2^m \ln\left(\frac{8}{\epsilon}\right) - 2\alpha \ln2.
    \label{eqn:N_cal}
\end{equation}
{To decrease $\epsilon$, each circuit is sampled a larger number of times, proportional to $\ln \left(\frac{1}{\epsilon} \right)$.}

If the iteration described above is used without fine-tuning, we take $\widetilde{\theta}_{\textrm{PI}}$ as the final estimate of $\theta$. {With a probability $1 - \epsilon$, this estimate differs from the true $\theta$ by a truncation, with an {error} equal to the maximum value of the bits not estimated: $\Delta \widetilde{\theta}_{\textrm{success}} = \frac{\pi}{2^m}$.} However, if the circuit with $N=2^i$ fails to identify the $(i+1)^{\textrm{th}}$ bit, this bit and the subsequent bits are incorrectly labelled. Therefore, $\widetilde{\theta}_{\textrm{PI}}$ differs from the true $\theta$ by up to twice the value of the $(i+1)^{\textrm{th}}$ bit: $\Delta \widetilde{\theta}_{\textrm{fail,i}} = \frac{\pi}{2^{i}}$. 
The final estimate from running the protocol once has a total {mean absolute error} bounded by the weighted sum of the values of $\Delta \widetilde{\theta}_{\textrm{success}}$ and $\Delta \widetilde{\theta}_{\textrm{fail,i}}$:
\begin{equation}
    \Delta \widetilde{\theta} = (1-\epsilon) \Delta \widetilde{\theta}_{\textrm{success}} + \sum_{i=0}^{m-1}\epsilon_i \Delta \widetilde{\theta}_{\textrm{fail,i}} = (1+m\epsilon)\frac{\pi}{2^{m}}.
\end{equation} 
In the asymptotic limit, where $m$ is large, a constant $\epsilon$ leads to $\Delta \widetilde{\theta} = \mathcal{O}(m2^{-m})=\mathcal{O}\left({\log N_{\textrm{PI}}}/{N_{\textrm{PI}}} \right)$. However, choosing $\epsilon$ to be a function of $m$ allows $\Delta \widetilde{\theta}$ to decrease inversely with a larger function of $N_{\textrm{PI}}$: the choice of $\epsilon = \mathcal{O}\left(\frac{1}{m} \right)$ results in optimal scaling, with $\Delta \widetilde{\theta} = \mathcal{O}(2^{-m})$ and $N_{\textrm{PI}} = \mathcal{O}(2^m \log m)$. The overall {mean absolute error} then scales as 
\begin{equation}
    \Delta \widetilde{\theta}=\mathcal{O}\left(\frac{\log (\log N_{\textrm{PI}})}{N_{\textrm{PI}}}\right).
\end{equation}
This scaling is a mere logarithm of a logarithm from the ideal Heisenberg Limit \footnote{When computing the root-mean-squared error, individual errors are squared before addition. Choosing $\epsilon = \mathcal{O}(2^{-m})$ optimizes the RMS error scaling as $\log N_{\textrm{PI}} / N_{\textrm{PI}}$.}. 

In our protocol, we combine this iterative point-identification step with the fine-tuning step, such that $N_{\textrm{max}}=2^m$, independent of $\epsilon$. Again, the total mean absolute error of the final estimate, $\Delta \widetilde{\theta}$, is the weighted sum of the standard deviation from success, $\sigma_{\widetilde{\theta}}$ [Eq. \eqref{eqn:cal_err}], and the error from failure of the point-identification, $\Delta \widetilde{\theta}_{\textrm{fail,i}}$:
\begin{equation}
    \Delta \widetilde{\theta} = (1-\epsilon)\sqrt{\frac{\ln2}{2^m \frac{\partial}{\partial m} \left( \alpha 2^m \ln\left(\frac{8}{\epsilon}\right) \right)}} + \frac{(m+1)\pi\epsilon}{2^m},
\end{equation}
in the asymptotic limit. The choice of $\epsilon = \mathcal{O} \left( m^{-3/2} \right)$ results in optimal scaling, with $N_{\textrm{tot}} = \mathcal{O}(2^m \log m)$ and
\begin{equation}
    \Delta \widetilde{\theta} = \mathcal{O} \left( \frac{\sqrt{\log (\log N_{\textrm{tot}})}}{N_{\textrm{tot}}} \right).
\label{eqn:It_cal_err_scale}
\end{equation}
This {error} scaling is a $\sqrt{\log (\log N_{\textrm{tot}})}$ improvement over our iterative point-identification alone \footnote{The root-mean-squared-error scaling improves similarly  to $\sqrt{\log N_{\textrm{tot}}} / N_{\textrm{tot}}$.}. Furthermore, in the simulations below, we see a significant reduction in the constant before the scaling.

\textit{Point identification using the QPE algorithm}.---The QPE algorithm employs  inverse Fourier transforms instead of MLEs to estimate $\theta$ \cite{Kitaev}. To gain a $b$-bits estimate of $T = \theta / 2\pi$ with an expected failure probability of $\epsilon$, $t = b + \lceil \log_{2}(2 + \frac{1}{2\epsilon}) \rceil$ qubits are manipulated with $N_{\textrm{tot}}=2^{t}-1$ applications of $\hat{U}(\theta)$ \cite{nielsen}. The maximum depth of the circuit is $2^{t-1}$ applications of $\hat{U}(\theta)$ plus a linear term, $\mathcal{O}(t)$, to apply the quantum Fourier transform. As such, the success probability, $1-\epsilon$, is only increased by increasing $N_{\textrm{tot}}$ and $N_{\textrm{max}}$: {$N_{\textrm{tot}} \propto \frac{1}{\epsilon}$}. The {error} scaling is the optimum Heisenberg Limit: $\Delta \widetilde{\theta} = \mathcal{O} \left( \frac{1}{N_{\textrm{tot}}} \right)$ \cite{Higgins}. Despite the optimal scaling of the QPE algorithm, the constant factor before the scaling causes inefficiency if the failure probability is low.
Furthermore, the large circuit requiring multi-qubit fully-entangled states create difficulty implementing the QPE algorithm with noisy intermediate-scale quantum hardware.

\begin{figure}[H]
    \centering
    \includegraphics[width = 8.2cm]{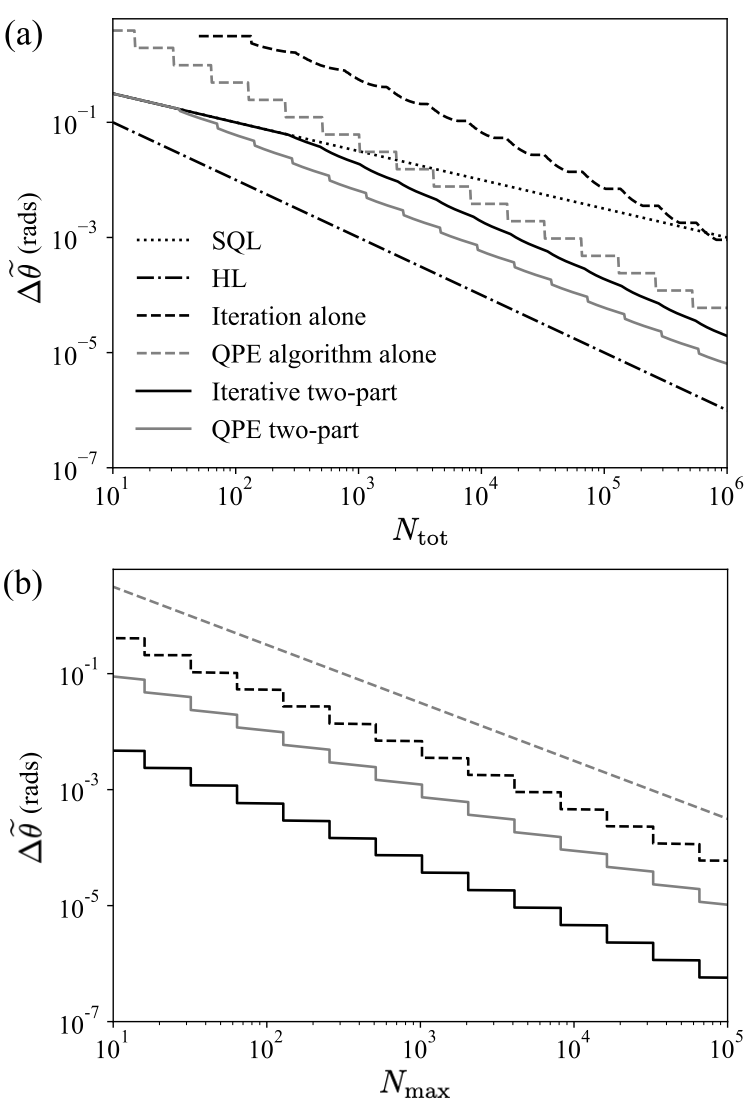}
    \caption{Numerical simulations of \textbf{(a)} $\Delta \widetilde{\theta}$ vs. $N_{\textrm{tot}}$ and \textbf{(b)} $\Delta \widetilde{\theta}$ vs. $N_{\textrm{max}}$ for the different protocols, with optimised $\epsilon$. Both plots use the same legend. SQL represents the Standard Quantum Limit and HL represents the Heisenberg Limit.}
    \label{fig:error_plots}
\end{figure}

When using the QPE algorithm in the point identification step of our two-step protocol, we choose $b=m+1$. Consequently, $\hat{U}(\theta)$ is applied in total $N_{\textrm{PI}} = 2^{m+1+\lceil \log_2(2+1/2\epsilon) \rceil}-1$ times. Equations \eqref{eqn:min_m} and \eqref{eqn:cal_err} lead to
\begin{equation}
\begin{split}
    N_{\textrm{tot}} &= 2^{m+2+\lceil \log_2(2+1/\epsilon) \rceil}-1=\mathcal{O}(2^m), \\
    \Delta \widetilde{\theta} &= \frac{1}{2^{m+1+\lceil \log_2(2+1/2\epsilon) \rceil/2}} = \mathcal{O} \left( \frac{1}{N_{\textrm{tot}}} \right).
\end{split}
\end{equation}
The {error} scaling still follows the Heisenberg Limit, but the constant before the scaling is smaller than the QPE algorithm alone. See Fig. \ref{fig:error_plots}(a). The largest circuit depth exceeds $N_{\textrm{max}} = 2^{m+\lceil \log_2(2+1/2\epsilon) \rceil}=\mathcal{O}(N_{\textrm{tot}})$ in the asymptotic limit. Circuits are thus deeper than the iterative techniques described above, and use currently impractical many-probe entanglement.

\textit{Simulations}.---In order to compare the performance of our two-step protocol to previous protocols, we provide numerical simulations. {We choose values of $m$ and $\epsilon$ that minimize the {error} of an estimate, $\Delta \widetilde{\theta}$, for a given value of $N_{\textrm{tot}}$. This value of $\Delta \widetilde{\theta}$ is then plotted against $N_{\textrm{tot}}$ [Fig. \ref{fig:error_plots}(a)] and $N_{\textrm{max}}$ [Fig. \ref{fig:error_plots}(b)], for each protocol. We also set $\alpha = 32$ to facilitate comparison to previous work \cite{RG}. Figure \ref{fig:error_plots}(a) shows that the iterative two-step protocol diverges from the Standard Quantum Limit at $\approx2300$ and $\approx7$ times smaller values of $N_{\textrm{tot}}$, compared to the iterative point-identification protocol alone and the QPE algorithm alone, respectively. That is, we reach quantum advantage with fewer applications of $\hat{U}(\theta)$ than other protocols. 
For all simulated values of $N_{\textrm{tot}}$, our protocol produces smaller values of $\Delta \widetilde{\theta}$, compared to a protocol that conducts iterative point identification alone, as well as compared to the QPE algorithm. This happens despite our protocol having a slightly inferior asymptotic scaling compared to the QPE algorithm.
Our two-step protocol with QPE-algorithm point identification achieves a lower {error} than the iterative two-step protocol for all simulated values of $N_{\textrm{tot}}$, but requires deeper circuits and entanglement. Figure \ref{fig:error_plots}(b) plots $\Delta \widetilde{\theta}$ as a function of $N_{\textrm{max}}$. The iterative two-step protocol is the most quantum-resource-efficient protocol, achieving the lowest {errors} with shallow circuits. To achieve an {error} below the Standard Quantum Limit for a given $N_{\textrm{tot}}$, the iterative method alone and the QPE algorithm alone require $N_{\textrm{max}} = 2048$ and $N_{\textrm{max}} = 1024$, respectively. Our two-step protocols achieve this for $N_{\textrm{max}}=2$ and  $N_{\textrm{max}}=8$, when iteration and the QPE algorithm is used for point identification, respectively.
}

\textit{Conclusion}.---We have proposed a new two-step phase-estimation protocol. In the first step, our protocol produces several contending precise estimates of an unknown phase, by sampling from a circuit with many applications of the unknown phase. Then, the protocol point identifies which estimate is in the correct parameter regime. This point identification involves independently sampling multiple circuits---each of which doubles in depth---in order to minimize the {error} of the final estimate of the phase with a set number of applications of the unitary operation, $N_{\textrm{tot}}$. 
For a given $N_{\textrm{tot}}$, our protocol achieves a lower {error} when compared to previous iterative protocols. Assymptotically, our protocol’s {mean absolute error} scales as $\mathcal{O} \left({\sqrt{\log (\log N_{\textrm{tot}})}} / {N_{\textrm{tot}}} \right)$, which is to be compared with a previously published, best, iterative scaling of $\mathcal{O} \left({\log N_{\textrm{tot}}}/{N_{\textrm{tot}}} \right)$ \cite{RG, Burgh}. 
Furthermore, when compared to the QPE algorithm, our protocol’s circuits are shallower, independent of the failure probability, and they do not require multi-qubit entanglement. 
Our protocol also achieves a lower {error} than the QPE algorithm for currently realistic values of $N_{\textrm{tot}}$, despite having a worse asymptotic scaling.
The achievement of a high precision with a shallow circuit suggests our protocol is more practical to implement in hardware-limited situations, such as noisy intermediate-scale quantum computers. 

\textit{Acknowledgements}.---The authors thank N. Mertig and W. Salmon for their support during this project. The authors also acknowledge support from Hitachi, Lars Hierta’s Memorial Foundation, and Girton College, Cambridge.

\bibliography{pp}

\end{document}